\documentclass{aa}
\usepackage{aabib99}
\usepackage{graphics}
\usepackage{times}

\chardef\isp="10 
\def\i{\'\isp}
\def\gsim {\ifmmode {\buildrel>\over\sim}		               
	\else {\lower.6ex\hbox{$\buildrel>\over\sim$}}\fi}
\def\lsim {\ifmmode {\buildrel<\over\sim}	   	               
	\else {\lower.6ex\hbox{$\buildrel<\over\sim$}}\fi}
\def\am {\ifmmode {^{\scriptscriptstyle\prime}}			       
	\else $^{\scriptscriptstyle\prime}$\fi}
\def\deg {\ifmmode^\circ\else$^\circ$\fi}			       
\def\eg{{\rm e.g.\/\ }}		
\def\raw {\ifmmode\rightarrow\else$\rightarrow$\fi}	
\def\uc{\rm J=1\raw0} 
\def\du{\rm J=2\raw1}					
\def \galpha {$\alpha $}      
\def\hdos {\hbox{${\rm H}_2$}}                    
\def\cmm#1{\ifmmode {\,{\rm cm^{-#1}}\;} 		              
	\else \hbox{$\,${\rm cm$^{\rm -#1}\;$}}\fi}
\def\menos{$-$}
\def\masmenos{\ifmmode {\pm} \else $\pm$ \fi}		               
\def\kmsns{~km~s$^{-1}$}
\def\kms{\ifmmode {{\rm \;km\;s^{-1}\;}}		    	      
        \else {\hbox{$\,${\rm km$\;$s$^{\rm -1}\;$}}}\fi}
\def\ie {i.\,e.}
\def\E#1 {\ifmmode {$\times$10^{#1}\;} 
	\else \hbox{$$\times$10^{#1}\;$}\fi}                                                
\def\apro{$\sim$}                                   
\def\gt{\ifmmode {>}\else{$>$}\fi}

\begin{document}
 
   \thesaurus{09        
              (09.03.1;  %
               09.13.2;);  %
               10.03.1;
               13 (13.19.1; %
               13.25.4) %
             }
   \title{A correlation between the SiO and the Fe 6.4 keV line emission from the Galactic center}
 
 
   \author{ J. Mart{\i}n-Pintado, P. de Vicente, 
  N. J. Rodr{\i}guez-Fern\'andez,  A. Fuente, and  P. Planesas}
 
   \offprints{ J. Mart{\i}n-Pintado}

   \institute{ Observatorio Astron\'omico Nacional (IGN), Campus        Universitario,  Apdo. 1143, E-28800 Alcal\'a de Henares, Spain \\}
 
   \date{Received, 2000; accepted, 2000}
 
   \authorrunning{Mart\i n-Pintado et al.}
   \maketitle
 
   \begin{abstract}
%
One of the most interesting results of the X-ray observations with
the ASCA satellite of the Galactic center (GC) is the  spatial  
distribution and the intensity of the Fe line at 6.4 
keV. Up to now the morphology
and the intensity of this line have been a puzzle. 
In this letter we present a map of the GC 
in the \uc\ line of SiO covering the same region than the ASCA observations.
The SiO emission is restricted to molecular clouds 
with radial velocity between  10 and 60 \kms
in the Sgr A and Sgr B complexes.
We find a correlation between the SiO morphology and  
the spatial distribution of the Fe 6.4 keV line, on the large scale and also within Sgr A and Sgr B.  The SiO abundance increases by a factor of $\gsim$20 in the regions with strong Fe 6.4 keV line. This indicates that the Fe 6.4 keV line mainly arises from molecular clouds with large gas phase abundance of refractory elements.
We discuss the implications of the correlation on
the origin of the hard X-rays, and the heating and the chemistry of the molecular clouds in the GC. 
 

\end{abstract} 


\section{Introduction}
The Galactic center (GC) is a strong source of diffuse X-ray emission in the 2-10 keV energy range  and in lines from several ions 
\cite{kawai88,sunyaev93,koyama96,koyama96,sidoli99}. Recently, 
the ASCA satellite mapped 
the X-ray emission from the GC \cite{koyama89}.
One of the most interesting results is the spatial 
distribution and the intensity of the iron K\galpha\ lines. The Fe K\galpha\ lines of highly 
ionized ions (He-like at $\sim$6.70 keV and H-like at $\sim$6.97 keV) arise from hot gas with temperatures of $\sim$9 keV. This emission is concentrated towards Sgr A-West 
and symmetrically distributed along the galactic plane with a spatial 
distribution similar to that of the radio continuum emission and the molecular 
clouds \cite{maeda96}. This is in sharp contrast with 
the K\galpha\ line from neutral or low ionized Fe atoms at $\sim$6.40 keV (hereafter Fe\deg\ line)
which shows emission only 
towards the Sgr A and Sgr B complexes \cite{koyama96}. 

The Fe\deg\ line emission is caused by fluorescence and appears when neutral 
cold molecular clouds are exposed to a strong source of hard X-rays. In 
X-ray irradiated molecular clouds like those in the GC, it is expected that 
the X-rays will influence  the heating, the ionization and the chemistry of 
these clouds (see \eg Hollenbach et al., 1997). It is well known that the 
physical conditions and the chemistry of the molecular clouds in the GC differ 
substantially from those in the galactic disk (see \eg 
Morris \& Serabyn, 1996).  
High gas kinetic temperature \cite{hutte93},  
and large abundance of SiO  are typical in the GC \cite{Minh92,martin97,hutte98}. 
The origin of these unusual characteristics is unclear, but it is believed to be due to strong shocks in 
the GC \cite{wilson82,martin97}. 
In this letter we present a correlation between the SiO radio emission and the Fe\deg\ line, suggesting that X-rays may play an important role in the heating and the chemistry of the GC molecular clouds. 


\begin{figure*}
\rotatebox{-90}{\includegraphics{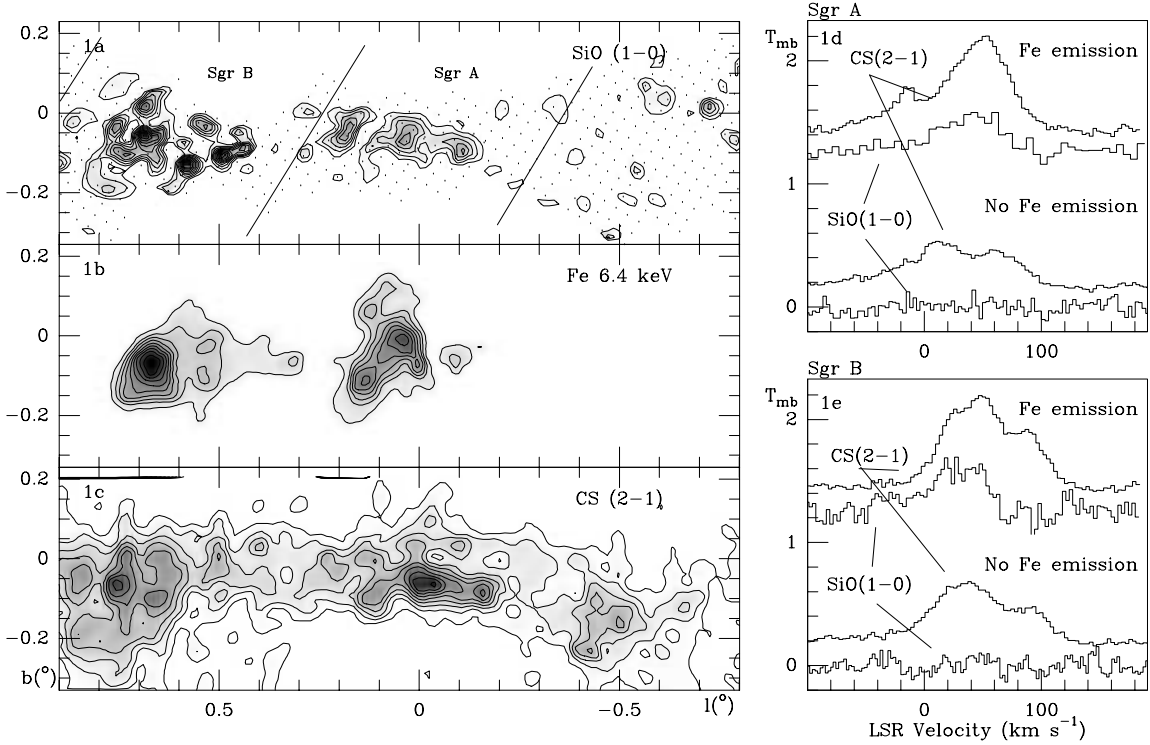}}
\caption{a-c) Spatial distribution of the integrated emission of the SiO \uc\ line, 
of the Fe\deg\ line (Koyama et al 1997) and of the CS \du\ line 
(Bally et al. 1987) 
in the GC. The velocity range used to calculate the integrated 
intensity for  SiO and  CS was  \menos100 to 100 \kmsns. The contour 
levels are from 10 to 100 K\kms in steps of 10 K\kms for the SiO map, and 15 to 150 K\kms 
in steps of 15 K\kms for the CS map. For the Fe\deg\ map the contour levels (in 10$^{-6}$ counts/sec/0.106 min$^2$) are 0.25 and 0.4 to 1 by 0.1. The dots in Fig. 1a show the location 
where the SiO spectra were taken. d-e) Comparison of the averaged line profiles of 
the \uc\ line of  SiO and of the \du\ line of CS for  Sgr A and Sgr B respectively 
for the regions with and without Fe\deg\ emission. 
The areas used to obtain the averaged spectra towards Sgr A and Sgr B  
are shown in Fig. 1a by parallel lines at constant Declination. The region with Fe\deg\ emission has 
been defined 
using a threshold intensity for this line of 0.25$\times$10$^{-6}$ 
counts/sec/0.106 min$^2$.}
\label{f:mapas}
\end{figure*}

\section{Observations and results}

The mapping of the \uc\ line of SiO was carried out with the 14-m radio telescope at 
the Centro Astron\'omico de Yebes in Guadalajara (Spain). The telescope characteristics, 
the receiver,  the backend, and the mapping procedure  have been described 
in Mart{\i}n-Pintado et al. \cite*{martin97}. The aperture efficiency 
of the telescope at low
elevation has been improved by a factor of 1.5 due to a shaped subreflector 
that corrects for
the large scale gravitational deformations in the main reflector \cite{garrido99}. 
The new SiO map, that contains the data by \cite{martin97},
has been extended to cover the region observed with the ASCA satellite.
The integrated intensity  map of the \uc\ line, obtained with 2\arcmin\ 
resolution, is shown in Fig. 1a. 

Figures 1b and 1c show the spatial distribution of the Fe\deg\ line 
\cite{koyama96} and 
of the integrated intensity of the \du\ line 
of CS \cite{bally87} obtained with angular resolutions 
of  2\arcmin\  and 3\arcmin\ respectively.
The spatial distribution of 
the SiO integrated intensity is different from that of CS. While the CS 
integrated intensity is relatively smooth with a moderate increase by  a factor of \lsim3 
towards the molecular cloud complexes Sgr A ($\ell\sim$0\deg), Sgr B ($\ell\sim$0.6\deg) 
and Sgr C ($\ell\sim$\menos0.5\deg), the SiO emission mainly arises from 
Sgr A and Sgr B. 
To show that the difference between the CS and the SiO emission cannot 
be due to dynamic range problems in the SiO map, the contour levels in 
the CS map have been chosen to be a factor 1.5 of those in the SiO map. 
Furthermore, the  SiO emission from Sgr A and Sgr B increases with 
respect to the surrounding clouds by a factor of $\gsim$8.  
The fine scale of the 
SiO and CS integrated line intensities 
also show important differences within the Sgr A and Sgr B complexes 
\cite{martin97}. 

\section {The correlation between the SiO abundance and the fluorescence Fe\deg\ line}

\begin{figure}
\rotatebox{-90}{\includegraphics{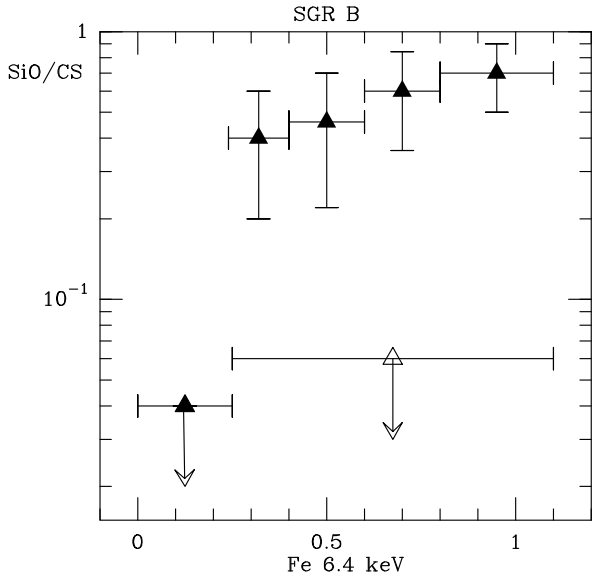}}
\caption{The ratio between the integrated line intensity of SiO and CS towards Sgr B as a function of the 
Fe\deg\ line intensity in units of 10$^{-6}$ counts/sec/0.106 min$^2$. The filled and the open triangles are for the 40 and the 90\kms clouds respectively.  The error bars shows the range of variation of the SiO/CS ratio within 
the region with strong Fe\deg\ emission.}
\label{f:correlacion}
\end{figure}

On large scales the morphology of the SiO emission shows a remarkable correlation with that of the Fe\deg\ line (Fig. 1b). Both emissions are only detected towards Sgr A and Sgr B, but not towards Sgr C.
Given the different requirements for the excitation 
of the SiO and the Fe\deg\ lines, departures from this correlation are also found. The main exception to the correlation is the presence of SiO emission towards the radio-Arc 
at $\ell\sim$0.18\deg\ where the Fe\deg\ line
is not detected. 
We do not find, however, the opposite situation, \ie\ strong Fe\deg\ without SiO emission.

There is also an overall correlation between the Fe\deg\  and the SiO 
emissions within Sgr A and Sgr B. This is illustrated in Fig. 1d and 1e, 
where we show  the spectra of the SiO and the CS emission averaged over
Sgr A and Sgr B (see Fig. 1a)
for the regions with and without
emission in the Fe 6.4 
keV line. While the CS line intensity is independent (within a factor of \lsim2) 
of the spatial distribution of the Fe\deg\ emission, the SiO line intensity changes by a factor of \gsim10 (\gsim20 for integrated intensities) between the regions with and without Fe\deg\ line emission. 

In the regions with emission in the Fe\deg\ line, we find that the line profiles of SiO and CS are different. The SiO emission mainly appears in the molecular clouds with radial velocities between 10 and 60 \kms (hereafter 40 \kms clouds). Towards Sgr B we do not detect SiO emission from the CS clouds with 
radial velocities between 70 and 110 \kms (hereafter 90\kms clouds).

Fig. 2 shows the ratio between the integrated line intensity of the 
SiO and the CS emission (hereafter the SiO/CS ratio) for the 40 \kms clouds (filled triangles) and for the 90 \kms clouds (open triangle) as a function of the intensity of the Fe\deg\ line for Sgr B. 
For the regions without or weak  (\lsim 0.25$\times$10$^{-6}$  counts/sec/0.106 min$^2$) Fe\deg\ line emission, we do not detect the SiO line and 
derive an upper limit to the SiO/CS ratio of 4$\times$10$^{-2}$. The SiO/CS ratio increases by more 
than a factor of 20 in the region with the strongest Fe\deg\ line emission. There is a weak increase of the SiO/CS ratio with the intensity of the Fe\deg\ line for intensities of \gt0.25$\times$10$^{-6}$ counts/sec/0.106 min$^2$. However, this change is marginal, since it is within the scatter in the SiO/CS ratios (see Fig. 2). 
The SiO emission is not detected for the 90\kms  clouds independently of the Fe\deg\ line intensity. The upper limit to the SiO/CS ratio  for these clouds in Sgr B is  6$\times$10$^{-2}$, similar to that derived for the molecular clouds without Fe\deg\ line emission.

The differences between the CS \du\ and the SiO \uc\ lines cannot be due to excitation effects  
since both transitions have similar rotational constants and similar critical 
densities. Opacity effects can also be ruled out since the two lines 
show similar optical depths \cite{hutte98}. Therefore, the  
difference between the CS and the SiO emission  is due to changes in the 
SiO abundance relative to that of CS \cite{martin97}. Changes in the CS 
abundance are unlikely since the spatial distribution of CS is similar to 
that of $^{13}$CO \cite{bally87}. 
For a CS fractional abundance of $\sim$5$\times$10$^{-9}$ and the typical physical conditions in the GC clouds (\hdos\ density of $\sim$5$\times$10$^{4}$  \cmm3 and a kinetic temperature of 40-200 K), the SiO abundance in the 40 \kms clouds for the region with Fe\deg\ line emission is  $\sim$10$^{-9}$.
The SiO abundance decreases to \lsim5$\times$10$^{-11}$ for the molecular clouds without Fe\deg\ line emission like Sgr C and for the 90\kms\ clouds in the regions with strong Fe\deg\ line in Sgr B. This low SiO abundance suggests that the bulk of the Fe\deg\ line arises only from the 40 \kmsns\ molecular clouds. Therefore, \apro60\% of the  mass in Sgr B, that with SiO emission, contributes to the Fe\deg\ line. This estimate is in agreement with the results obtained from the modeling of the Fe\deg\ line 
intensity in Sgr B which requires that 50\% of the  mass contributes to the Fe\deg\ line \cite{murakami99}. In summary, the Fe\deg\ line emission in the GC arises from the molecular clouds with large gas phase abundance of refractory elements like SiO.  

\section {Discussion}

The relationship between the SiO abundance
and the presence of the Fe\deg\ line opens the possibility to understand the origin of SiO in gas phase and that of the Fe\deg\ line in the GC. 
The molecular clouds with emission in the Fe\deg\ line always seem to have enhanced abundance of SiO in gas phase. The simplest explanation could be that this association is related to a metal abundance larger than solar in some  molecular clouds as indicated by modeling of the 
Fe\deg\ line intensity in Sgr B \cite{murakami99}.
Metal enriched molecular clouds in the GC could also have 
more Si in gas phase which would be completely converted into SiO \cite{herbst89}. 
However,
changes in the cosmic abundance from cloud to cloud  seem to be unlikely since the metallicity in the hot diffuse component is rather homogeneous over the 
1\deg$\times$1\deg region around the GC \cite{maeda96}. 

A common origin for the X-rays and the large SiO abundance in gas phase and/or a peculiar chemistry induced by X-rays could both explain the correlation. 
The origin of the X-rays producing the Fe\deg\ line in the GC is unclear 
\cite{koyama96,sunyaev98}. The relatively large abundance of SiO and its 
spatial distribution in the GC clouds can be explained by an increase of Si 
or SiO in gas phase due to grain processing by shocks \cite{martin97,hutte98}. 
It is possible that the sources driving the strong shocks responsible for 
the grain destruction 
also generate the hard X-ray 
emission  which excite the Fe\deg\ line. This could be the case for  Sgr A where an explosion in Sgr A East, which is now 
expanding inside the 50 \kms cloud, has been proposed as the origin of the high-energy 
activity in this region \cite{yusef97}.

For the Sgr B cloud, it has been argued that  
the Fe\deg\  line cannot be excited by the hot plasma since the observed X-ray luminosity 
is one order of magnitude smaller than the required to account for that observed in the Fe\deg\ line. 
Therefore it has been proposed that 
this complex is a X-ray reflection nebula  illuminated by a time variable 
source(s) of hard X-rays located outside the neutral cool material \cite{koyama96}. 
It is possible that a burst of X-rays that occurred  hundreds of years ago from the Sgr A region 
(likely from Sgr A*)  now irradiates the Sgr B  molecular complex 
\cite{koyama96,sunyaev98}. 
In the X-ray reflection nebula scenario, the heating, ionization and the 
chemistry of the molecular clouds will be influenced by the illuminating 
X-rays \cite{maloney96}. Molecular clouds irradiated by X-rays contain 
regions of high temperature where reactions with activation barriers could 
contribute to molecule formation \cite{neufeld94}. Then, the SiO abundance 
in the hot regions could be enhanced if Si is in gas phase. If silicate 
grains smaller than 10 \AA\ are present, X-rays can evaporate these 
dust grains \cite{voit91} providing the small fraction of Si in gas 
phase (0.1\%) required to explain the SiO abundance in the GC molecular clouds.

It is interesting to note that 
the peak of the Fe\deg\ line in Sgr B  coincides with a 
large concentration of hot expanding molecular shells \cite{martin99}. 
These shells are thought to be produced 
by wind-blown bubbles driven by massive evolved stars.  Massive evolved stars in the  supernova 
stage show both the Fe\deg\ line (see \eg Kinugasa et al. 1999) and a large 
abundance of  SiO \cite{ziurys89}. It is possible that any
of the sources driving the hot expanding
molecular shells has undergone about 30 years ago a flare in hard X-rays giving rise to the observed Fe\deg\ line.  

\section {Conclusions and perspectives}

We have presented a new map of the SiO emission from the inner 200 pc of the Galaxy. We find a correlation between the spatial distribution of the SiO emission and that of the fluorescence line of Fe 
at 6.4 keV. The Fe\deg\ line is only found in the molecular clouds where the  
SiO abundance increases by more than a factor of 20 with respect to the molecular cloud complexes without Fe\deg\ line emission. 
The origin of this association is so far unknown. X-ray sources driving strong shocks or X-ray reflection nebula are possible explanations. Obviously, the combination  of  high angular resolution maps of the hard X-ray emission obtained with XMM and Chandra with those
of the molecular emission from refractory elements  will provide important clues about the origin of the X-ray activity 
and the unusual properties of the molecular clouds in the GC. 

\acknowledgements{We thank Dr. J. Bally for kindly providing the CS data and Drs. T.L. Wilson and 
R. Gaume for critical reading of the manuscript. This work has been 
partially supported by the Spanish DGICYT under grants PB96-104 and 1FD97-1442. NJR-R has been supported by the Consejer{\i}a
de Educaci\'on y Cultura de la Comunidad de Madrid.}

\bibliography{aamnem99,xrays-sio}
\bibliographystyle{aabib99}

\end{document}